\documentclass[3p,times]{elsarticle}

\usepackage{ecrc}
\usepackage{epstopdf}
\volume{00}
\firstpage{1}
\journalname{Nuclear Physics A}
\runauth{O.~Kaczmarek}

\jid{nupha}

\jnltitlelogo{Nuclear Physics A}

\usepackage{graphicx}
\usepackage{amsmath,amssymb}

\input pix.sty

\def\undertilde#1{\mathop{\vtop{\ialign{##\cr$\textstyle{#1}$\cr%
\noalign{\kern1pt\nointerlineskip}\hfil$\mathchar"0365$\hfil\cr}}}}
\def\wideundertilde#1{\mathop{\vtop{\ialign{##\cr$\textstyle{#1}$\cr%
\noalign{\kern1pt\nointerlineskip}\hfil$\mathchar"0367$\hfil\cr}}}}

\def\lsi{\raise0.3ex\hbox{$<$\kern-0.75em\raise-1.1ex\hbox{$\sim$}}}
\def\gsi{\raise0.3ex\hbox{$>$\kern-0.75em\raise-1.1ex\hbox{$\sim$}}}

\newcommand{\rmii}[1]{{\mbox{\tiny\rm{#1}}}}

\newcommand{\re}{\mathop{\mbox{Re}}}

\newcommand{\Tint}[1]{{\hbox{$\sum$}\!\!\!\!\!\!\!\int\,}_{\!\!\!\!\raise-0.9ex\hbox{$\scriptstyle{#1}$}}}
\newcommand{\Tinti}[1]{{{\Sigma}\!\!\!\!\raise0.3ex\hbox{$\int$}_\rmii{${#1}$}}}

\newcommand{\bi}{\begin{itemize}}
\newcommand{\ei}{\end{itemize}}
\newcommand{\hide}[1]{ }

\begin{document}

\begin{frontmatter}

\title{Continuum estimate of the heavy quark\\momentum diffusion coefficient $\kappa$}

\author{O.~Kaczmarek\fnref{col1}}
\ead{okacz@physik.uni-bielefeld.de}
\fntext[col1] {In collaboration with A.~Francis, M.~Laine, M.~M\"uller, T.~Neuhaus and H.~Ohno.}
\address{Fakult\"at f\"ur Physik, Universit\"at Bielefeld, D-33615 Bielefeld, Germany}

\begin{abstract}
Among quantities playing a central role in the theoretical interpretation of
heavy ion collision experiments at RHIC and LHC are so-called transport
coefficients. Out of those heavy quark diffusion coefficients play an important
role e.g. for the analysis of the quenching of jets containing
$c$ or $b$ quarks ($D$ or $B$ mesons) as observed at RHIC and LHC \cite{Beraudo:2014iva}.

We report on a lattice investigation of heavy quark momentum diffusion within
pure SU(3) plasma above the deconfinement transition with the quarks treated
to leading order in the heavy mass expansion. We measure the relevant
``colour-electric'' Euclidean correlator and based on several lattice spacings
perform the continuum extrapolation. This extends our previous studies
\cite{measure,Francis:2013cva}
progressing towards a removal of lattice artifacts and a physical
interpretation of the results.

We find that the correlation function clearly exceeds its
perturbative counterpart which suggests that at temperatures just above the
critical one, non-perturbative interactions felt by the heavy quarks are
stronger than within the weak-coupling expansion. 
Using an Ansatz for the spectral function which includes NNLO perturbative
contributions we were able to determine, for the first time, a continuum
estimate for the heavy quark momentum diffusion coefficient.
\end{abstract}


\end{frontmatter}

\section{Introduction}

In the relevant temperature regime for heavy ion experiments, perturbative
calculations of diffusion coefficients suffer from poor convergence
\cite{Moore:2004tg,CaronHuot:2007gq} and non-perturbative contributions may
become relevant.
In this work we determine
continuum results for the colour-electric correlation function from lattice QCD
calculations in the deconfined phase at a temperature of around $1.4~T_c$ and
use an Ansatz for the corresponding spectral function to determine a continuum
estimate of the heavy
quark momentum diffusion coefficient.

\begin{figure}
\begin{center}
\includegraphics*[width=10cm]{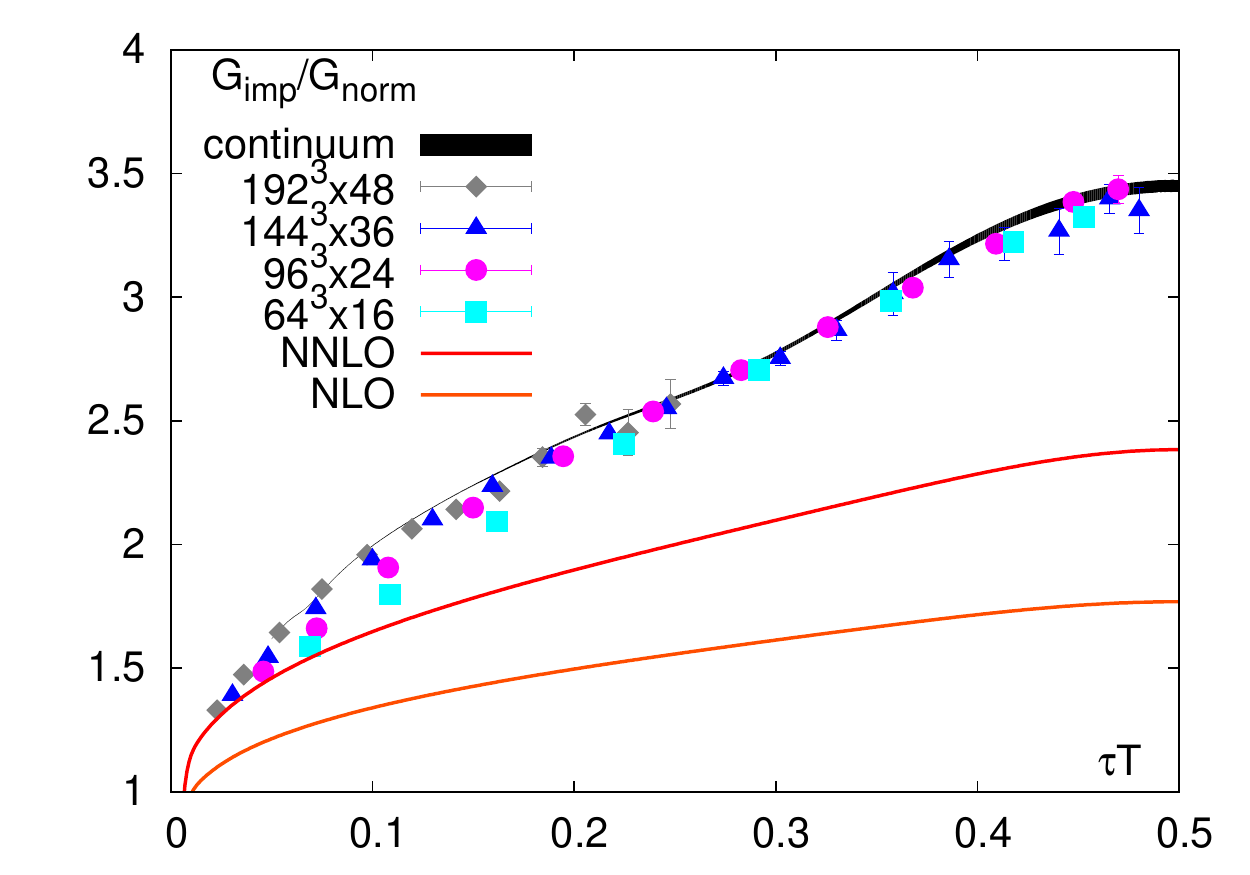}
\caption{
Lattice results for the colour-electric correlation function together
with the continuum extrapolated correlator. Also shown are results from a NLO and NNLO
perturbative calculation. 
}
\label{fig:generic}
\end{center}
\end{figure}

%

\begin{figure}[h]
\begin{center}
  \begin{minipage}[t]{8.5cm}
    \vspace*{-1cm}
    \includegraphics*[width=7.5cm]{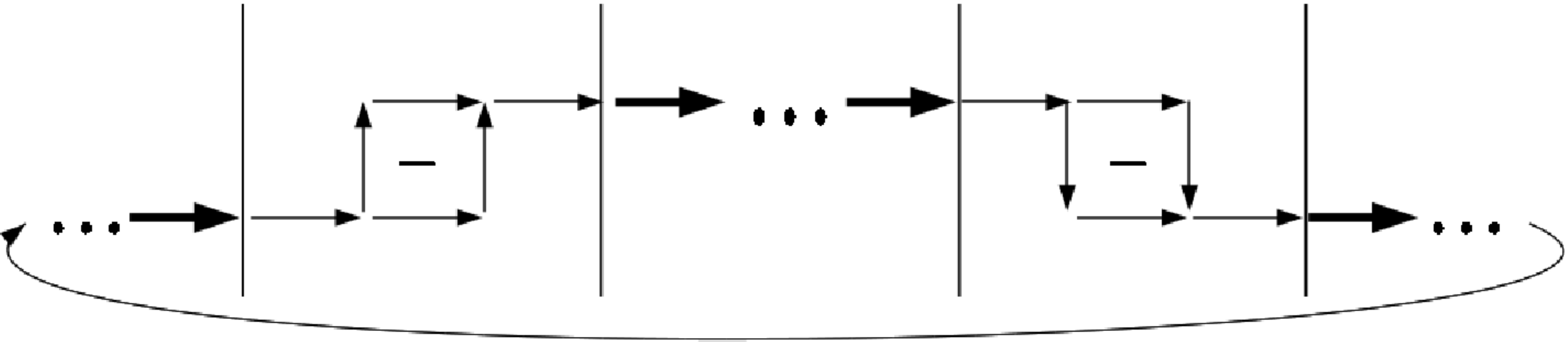}
  \end{minipage}
\hspace*{7mm}
  \begin{minipage}[t]{7.0cm}
    \small
    \begin{tabular}{|c|c|c|c|c|c|}
    \hline
       $\beta$ & $N_\tau$ & $N_s$ & $N_{\mathrm{conf}}$ & $N_{\mathrm{stat}}$
       & $r_0 T$  \\
    \hline\hline
     6.872 & 16 & ~64 & 100 &  1000 &   1.111 \\
     7.192 & 24 & ~96 & 160 &  1000 &   1.077 \\
     7.544 & 36 & 144 & 563 &  1000 &   1.068 \\
     7.793 & 48 & 192 & 223 &  1000 &   1.055 \\
    \hline
    \end{tabular}
  \end{minipage}
\end{center}
\caption[a]{
Left:
 Fat links, thin links and electric fields along the time direction
 (cf.\ the text).
Right:
 Run parameters.
 The values of $r_0 T$ were estimated in \cite{Francis:2013cva}.
}
\label{fig:multilevel-linkintegrated}
\end{figure}

Using Heavy Quark Effective Theory (HQET), the
propagation of a heavy colour charged quark and its response to a coulored Lorentz
force can be related through linear response theory to a ``colour-electric
correlator''~\cite{pos4,eucl},  
\be
 G_\rmi{\,E}(\tau) \equiv - \fr13 \sum_{i=1}^3 
 \frac{
  \Bigl\langle
   \re\tr \Bigl[
      U(\fr{1}{T};\tau) \, gE_i(\tau,\vec{0}) \, U(\tau;0) \, gE_i(0,\vec{0})
   \Bigr] 
  \Bigr\rangle
 }{
 \Bigl\langle
   \re\tr [U(\fr{1}{T};0)] 
 \Bigr\rangle
 }
 \;, \la{GE_final}
\ee
where $gE_i$ denotes the  
colour-electric field, $T$ the temperature, and $U(\tau_2;\tau_1)$
a Wilson line in the Euclidean time direction. 
A discretized version of this correlator 
is shown in Fig.~\ref{fig:multilevel-linkintegrated}.
The momentum diffusion coefficient can be obtained from the slope of the
spectral function in the low frequency limit, 
\begin{eqnarray}
\kappa/T^3 = \lim_{\omega\rightarrow 0} \frac{2 \rho_E(\omega)}{\omega T^2},
\end{eqnarray}
where $\rho_E(\omega)$ is the corresponding spectral function that is related
to the operator (\ref{GE_final}).
In the non-relativistic limit (i.e. for a heavy quark mass $M\gg \pi T$)
$\kappa$ is related to the diffusion coefficient $D=2 T^2 / \kappa$.

\section{Lattice determination of the heavy quark momentum diffusion coefficient $\kappa$}
We have performed quenched lattice QCD calculations of the discretized version
of the correlation function (\ref{GE_final}) (see
Fig.~\ref{fig:multilevel-linkintegrated}~(left))
using the standard Wilson gauge action on 4 different lattices at a temperature
around 1.4~$T_c$ listed in Tab.~\ref{fig:multilevel-linkintegrated}. As
the correlation
function decreases rapidly with $\tau$ and suffers from a weak signal-to-noise
ratio, noise reduction techniques are important to obtain a good signal,
especially at larger separations $\tau T\sim \frac{1}{2}$.  
We used multi-level updates \cite{lw,shear} for the part of the 
operator that includes the electric field insertions and 
link-integration (``PPR'') \cite{ppr,fr} for the straight lines between
them (the ``fat links'' in Fig.~\ref{fig:multilevel-linkintegrated}). 
As demonstrated in \cite{Francis:2013cva}, these techniques suffice to yield a
good signal.
Furthermore we use a tree-level improvement \cite{shear,rs} to reduce cut-off
effects and a NLO perturbative renormalization factor $Z_{pert}(\beta)$.
In Fig.~\ref{fig:generic} the lattice results for $G_{imp}(\tau)$ are shown,
normalized to 
\begin{displaymath}
 G_\rmi{norm}(\tau T)
 \; \equiv \;
 \pi^2 T^4 \left[
 \frac{\cos^2(\pi \tau T)}{\sin^4(\pi \tau T)}
 +\frac{1}{3\sin^2(\pi \tau T)} \right]
 \;.
\end{displaymath}
Although cut-off effects are visible at small separations and the results
become more noisy at large distances on the finer lattices, the results on the four lattices allow
for a controlled continuum extrapolation down to distances around $\tau T\sim
0.05$.
To obtain the continuum estimate of the correlation function, we perform b-spline
interpolations for each lattice and at fixed $\tau T$ extrapolate the
correlator in $1/N_\tau^2$. The result of this continuum extrapolation is shown in
Fig.~\ref{fig:generic} as a black solid band. Also shown are the NLO
\cite{rhoE} and NNLO correlation functions.

\begin{figure}
\begin{center}
\includegraphics*[width=10cm]{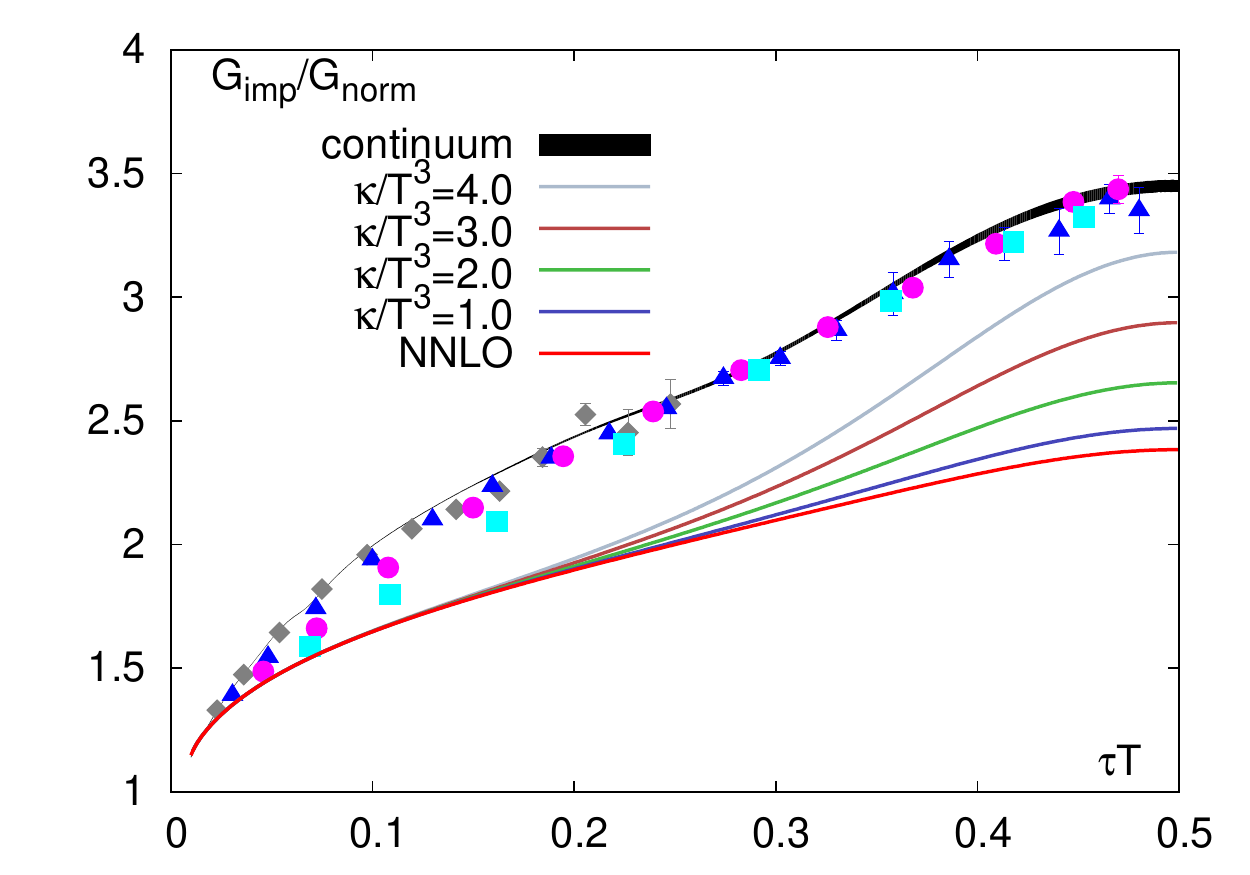}
\caption{
$G_{model1}$ for different values of the diffusion coefficient $\kappa$.
}
\label{fig:nnlotrans}
\end{center}
\end{figure}

\begin{figure}
\begin{center}
\includegraphics*[width=10cm]{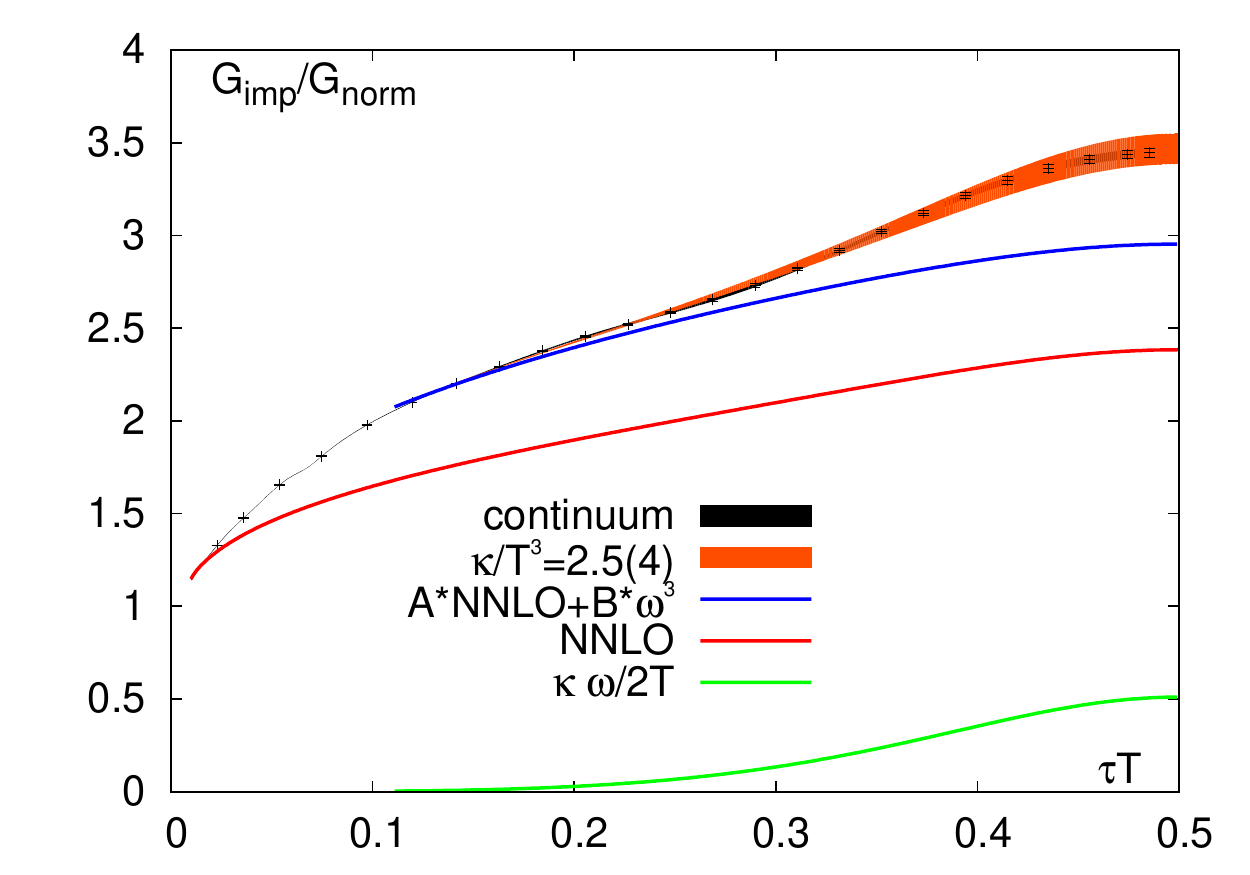}
\caption{
Result of the fit of $G_{model2}$ to the continuum extrapolated
correlator and the different contributions to the spectral function
Ansatz $\rho_{model2}$.
}
\label{fig:diff_fit2}
\end{center}
\end{figure}

In contrast to correlation functions of conserved currents where a Breit-Wigner
like transport peak is expected, studies of the
momentum diffusion operator in classical lattice gauge theory 
\cite{Laine:2009dd} and $\mathcal{N}=4$ Super-Yang-Mills in the large $N_c$ limit
\cite{pos4,Gubser:2006nz} suggest a rather flat behavior of
$\rho_E(\omega)/\omega$ in the small frequency limit.\\
As a first model spectral function we use
\begin{eqnarray}
\rho_{model1}(\omega) = \max\left\{ \rho_{NNLO}(\omega), \frac{\omega\kappa}{2 T}\right\},
\end{eqnarray}
and vary the momentum diffusion coefficient $\kappa$ in a range of $T^3$ and
$4~T^3$. From the results shown in Fig.~\ref{fig:nnlotrans} it is obvious that
this alone can not describe the data well. To allow for more 
contribution at intermediate frequencies region we use the Ansatz
\begin{eqnarray}
\rho_{model2}(\omega) = \max\left\{ A \rho_{NNLO}(\omega) + B \omega^3, \frac{\omega\kappa}{2 T}\right\},
\label{eq:model2}
\end{eqnarray}
containing three parameters $A$, $B$ and $\kappa$, and fit the corresponding Euclidean correlator
\begin{eqnarray} 
 G_{model2}(\tau) =
 \int_0^\infty
 \frac{{\rm d}\omega}{\pi} \rho_{model2}(\omega)
 \frac{\cosh \left(\frac{1}{2} - \tau T \right)\frac{\omega}{T} }
 {\sinh\frac{\omega}{2 T}} 
 \;
\end{eqnarray} 
in the range [0.1:0.5] to the continuum extrapolated data and obtain a
continuum estimate for the momentum diffusion coefficient 
\begin{eqnarray}
\kappa/T^3 = 2.5(4),
\end{eqnarray}
where the error was estimated by varying the fit-range and varying $\kappa$ such
that the $\chi^2/dof$ is of order unity. The result of this fit gives a good 
description of the data as shown in Fig.~\ref{fig:diff_fit2}. Also shown are
the different contributions from the model spectral function.\\
The result is compatible with previous estimates on finite lattices \cite{mumbai}
and predictions from the T-matrix approach~\cite{Riek:2010fk}.
Converted to the diffusion coefficient $D$ our estimate is larger than 
a lattice determination of $D$ for charm quarks on finite lattices
\cite{Ding:2012sp}. It remains to be seen how this determination will depend on
the quark mass and on a continuum limit. Work in this direction is in progress
\cite{Ohno:2013rka}.

%

\section*{Acknowledgment}
This work has been supported
in part by the
DFG under grant GRK 881
and by the European Union through I3HP and ITN STRONGnet,
Numerical calculations have been
performed using JARA-HPC resources at the RWTH Aachen Compute Cluster, JUQUEEN
at the JSC J\"ulich, the OCuLUS Cluster at the Paderborn Center for Parallel
Computing and the Bielefeld GPU Cluster.

\end{document}